\documentclass{aa}
\usepackage{natbib}
\bibpunct{(}{)}{;}{a}{}{,}
\usepackage{graphicx}
\usepackage{times}

\begin{document}

\title{Wide field weak lensing observations of A1835 and A2204}

\author{D. Clowe \inst{1,2} \and P. Schneider \inst{1,2} }

\offprints{D. Clowe}

\institute{Institut f\"ur Astrophysik und Extraterrestrische Forschung der 
Universit\"at Bonn, Auf dem H\"ugel 71, 53121 Bonn, Germany \and 
Max-Planck-Institut f\"ur Astrophysik, Karl-Schwarzschild-Str. 1, 85748
Garching, Germany}

\date{Received 19 April 2002 / Accepted 05 August 2002}

\abstract{
We present mass reconstructions from weak lensing for the galaxy clusters
A1835 and A2204 over
$34\arcmin \times 34\arcmin$ fields using data from the ESO/MPG Wide Field
Imager.  Using a background galaxy population of $22<R<25.5$ we detect the
gravitational shear of A1835 at $8.8\sigma$ significance, and obtain best-fit
mass profiles of $\sigma_v=1233^{+66}_{-70}$ km/s for a singular isothermal sphere
model and $r_{200}=1550 h^{-1}$ kpc, $c=2.96$ for a `universal' CDM profile.
Using a color-selected background galaxy population of $22<R<25.8$ we detect
the gravitational shear of A2204 at $7.2\sigma$ significance, and obtain
best-fit mass profiles of $\sigma_v=1035^{+65}_{-71}$ km/s for a SIS model
and $r_{200}=1310 h^{-1}$ km/s, $c=6.3$ for a `universal' CDM profile.
The gravitational shear at distances greater than $10\arcmin $ is significantly
detected for both clusters.  The best fit weak lensing cluster masses agree 
well with both X-ray and dynamical mass measurements, although the
central concentration of A1835 is much lower in the weak lensing mass profile
than that measured by recent Chandra results.  We suggest that this lower 
concentration is most likely a combination of contamination of the 'background'
galaxy population with cluster dwarf galaxies and the effect of a prolate
or tri-axial cluster core with the major axis lying near the plane of the sky.
We also detect a number of additional structures at moderate significance,
some of which appear to be sub-haloes associated with the clusters.
}

\maketitle

\section{Introduction}
In recent years, cosmological N-body simulations have been used to obtain
theoretical mass profiles for massive clusters of galaxies (\citealt{NA97.6},
hereafter NFW; \citealt{MO99.5}; \citealt{JI00.1}).  While
there is some discrepancy between the models for the mass profile at small 
radii ($\la 100$ kpc), at large radii ($\ga 1$ Mpc) they agree that the mass 
density should decrease as $r^{-3}$.  This is a much steeper slope at large 
radii than the singular isothermal sphere (hereafter SIS) model, which falls
as $r^{-2}$.  Measuring the mass profiles of clusters at large radii,
therefore, should provide an excellent test of the predictions from
N-body simulations.

Weak gravitational lensing, in which one determines the mass of an object
by measuring the shear induced in background galaxies by the gravitational
potential of the object, is a powerful tool for determining the mass and mass
profile of clusters of galaxies.  Because the gravitational shear depends
linearly on the two-dimensional mass surface density, it can be used to measure
mass profiles to much larger radii than X-ray observations, for which
the emissivity scales as the square of the baryonic mass density.  Further, the
resulting mass measurements have no dependence on the dynamical state of the
cluster and are not influenced by shocks caused by infalling material.
To date, the primary limitation on the maximum radius to which the cluster
mass profiles has been measured with weak lensing ($\sim 400-600 h^{-1}$ kpc) has 
been the size of the detector and the fact that the depth of exposure needed 
to obtain a sufficient number density of background galaxies for shear 
analysis made large-area mosaics impractical in telescope time.

With the advent of wide-field mosaic CCD detectors \citep{LU98.1}, it
is now possible to obtain a weak lensing signal to large radii ($\ga 2 h^{-1}$
Mpc) on moderate redshift clusters in a few hours of telescope time.
The mass profile for A1689 out to $2 h^{-1}$ Mpc was measured by 
\citet[hereafter CS01]{CL01.1}.  
Here we present the mass profiles to similar radii
for two additional massive clusters, A1835 and A2204.  In Sect.~2 we discuss
the image reduction and object catalog generation.  We analyze the weak
lensing signal for A1835 in Sect.~3 and for A2204 in Sect.~4.  In Sect.~5
we present out conclusions.  Unless otherwise stated we assume a 
$\Omega _\mathrm{m} = 0.3, \Lambda = 0.7$ cosmology, $H_0 = 100 h$ km/s/Mpc,
and that the mean redshift of faint galaxies is at $z_\mathrm{bg}=1$.

\section{Observations and Data Reduction}
Imaging on the fields was performed on the nights of May 29-30, 
2000, with the Wide Field Imager (WFI) on the ESO/MPG 2.2m telescope 
in La Silla, Chile.  In total, twelve 900 second
exposures in $R$ were obtained on A1835 on May 30, twelve 900 second exposures
in $R$ were obtained on A2204 on May 29, and three 600 second exposures in $B$
and $V$ were obtained on A2204 on May 30.
The night of May 29 was photometric, while the night of May 30 was
non-photometric and large changes in the brightness of objects can be seen
in consecutive images.  The images were taken with a dither pattern which
filled in the gaps between the chips in the CCD mosaic, resulting in the
final $R$-band images having $\sim 82\%$ of their area receiving the full
exposure time and the rest receiving varying amounts of lesser exposure times
as the area was in the chip gap or out of the field of view for one or more 
of the images.  The $V$ images were taken with large enough
offsets to fill in the chip gaps, but the final image has $\sim 12\% $ of the 
area covered by a single input image.  Because one of the three $B$ images is
unusable due to severe extinction, the coadded $B$ image has $\sim 24\% $ of 
the area covered by just one input image and $\sim 3\% $ of the area is blank.

\subsection{Image Reduction}

The image reduction technique for the $R$-band images is identical to that
given in CS01.  Treating each chip as its own separate camera, we de-biased
the images from a master bias taken at the beginning and end of each night,
and corrected for bias drift during the night by subtracting a linear fit to
the residual of the overscan strip.  The images were then flattened by a 
9th by 17th order two-dimensional polynomial fit to the twilight flats.  The
fit to the twilight flat was performed in order to remove the sky interference 
fringes present in the $R$-band data.  All of the long exposure $R$-band 
images from each night were then normalized with the mode sky value of the
image and medianed together.  The resulting night sky super-flat was also fit with
a 9th by 17th order polynomial, and the images were flattened with this
polynomial.  The flattened images were re-normalized and medianed again to
produce an image of the fringes, which have a peak-to-peak amplitude of
$\sim 10\%$, which was then scaled to each images' mode
sky value and subtracted.  As this technique tends to incorrectly subtract 
regions of low quantum efficiency, such as dust spots, such regions were 
masked by hand.  As was mentioned in CS01, due to the large number of input
images and the dithering between the images, the lensing results from this 
technique are indistinguishable from those produced by simply dividing by
the nightsky flatfield with the fringes left in place.  This technique,
however, should more accurately preserve the photometry of the objects
than by dividing by the fringes, particularly around the edge of the image
where few input images were averaged.  { We note that neither of these
methods is in principle correct, but lacking the means to separate the 
fringes from the
flat-field for this data set, these two methods were chosen to determine
if the miscorrection of the fringes or pixel variations could produce a
noticable systematic error in the lensing results.  Because the lensing
data from the two methods are statistically indistinguishable, we therefore
conclude that any systematic error produced is much smaller than the noise
in the data.}  The $B$ and $V$-band images were
flattened with a medianed flat without the fitting procedure above as the
fringes are present only in the $R$-band data.

The sky level in each image was determined by detecting minima in a smoothed
image.  An image containing only the minima was then smoothed by a 128 pixel,
$\approx 30\farcs7$, FWHM
Gaussian and divided by a similarly smoothed image containing the number
of minima (1 or 0) in each pixel.  This produced an image of the sky smoothed
on the $\approx 30\arcsec$ scale, which was then subtracted from the original
image.  This process was repeated on the sky-subtracted image using 96, 64,
48, and 32 pixel FWHM Gaussians.  The first sky-subtracted image which only
subtracts sky variations on the $\approx 30\arcsec$ scale was subsequently
used for the photometric measurements of cluster galaxies, while the final
sky-subtracted image which removed sky variations on a $\approx 7\farcs5$
scale, along with extended wings of bright stars and many of the
larger galaxies,  was used for the weak lensing measurements.  The $B$ and
$V$-band images had only the first sky-subtraction performed, as they are used
only for photometry.  The sky-fitting
routine performed in CS01 could not be used on these fields due to the presence
of several large stellar reflection rings in each field.

Stellar reflection rings are quite predominant on the WFI images.  Every
saturated star with blooming in the core has two reflection rings, roughly
$48\arcsec$ in diameter, centered a few arcseconds from the star, radially
away from the center of the image with the offset increasing with distance
from the center of the image, with typically $2\arcsec$ offset between the
two rings.  Stars which are bright enough to have the blooming extend beyond
the stellar core also have a visible third fainter reflection ring, 
$\sim 92\arcsec$ in diameter, which is centered about twice as far from the 
star and in the same radial direction as the two smaller rings.  Exceptionally
bright stars, however, reveal far more reflection rings.  A star in the SW
corner of the A1835 image, listed as $V=7.4$ in the USNO-A2.0 catalog \citep{USNO}, has three
additional reflection rings: a $318\arcsec$ diameter ring centered $140\arcsec$
NE of the star, a $232\arcsec$ diameter ring centered $287\arcsec$ NE of the
star, and a very faint ring $\sim 10\arcmin$ in diameter centered 
$\sim 1\arcmin$ NNE of the star.  A star in the WSW edge of the A2204 image,
$V=7.8$ in the USNO-A2.0 catalog, has two observable reflection rings, similar in
size and offset to the first two given above.  A $V=5.6$ star located only
a few arcminutes SW of the A2204 cluster near the center of the image also has
the same three additional rings, although the first is centered roughly on the
star, the second about $10\arcsec$ NE of the star, and the larger third ring
centered $\sim 4\arcmin$ SSW of the star.  In addition, there appear to be
a number of smaller reflection rings, $\sim 3\arcmin$ in diameter, located
$\sim 7\arcmin$ SSW of this star which overlap each other sufficiently to
make it impossible to determine an accurate count.  

The sky-subtraction process above was able to remove the largest reflection
ring such that we were confident of being able to accurately measure both
magnitudes and second moments of the surface brightness for objects in this
ring.  For the other two larger rings, the sky-subtraction software was
unable to remove the edges of the ring, including the spider shadows, so 
objects in these regions had their magnitudes and colors measured, but were
excluded from the lensing analysis.  The regions containing the three 
brighter rings found around all the stars were excluded from all subsequent
analysis, both lensing and photometric.

The next step is to simultaneously combine the eight CCD images into a single 
image, move it to a common reference frame from all the images of a given 
field, and remove any distortion in the image introduced by the telescope 
optics.  To do this we assumed that each CCD of the mosaic could be mapped
onto a common detector plane using only a linear shift in the $x$ and $y$
directions and a rotation angle in the $x$-$y$ plane.  We also assume that
these mappings are the same for all the data.  In doing so we are assuming
the CCDs are aligned sufficiently well vertically that there are no changes
in depth at which the focal plane is sampled across the chip boundaries and
that the CCDs do not move relative to each other when the instrument is
subjected to thermal variations or flexure during the night.  These assumptions
are verified later by the inability to detect a change in the point spread
function (PSF) across the chip boundaries.

We then used a bi-cubic polynomial to map the detector plane for each exposure
to a common reference frame for each field.  The parameters for the two
coordinate mappings were determined simultaneously by minimizing the
positional offsets of bright but unsaturated stars among the images of the
same field and with the USNO-A2.0 star catalog.  A downhill simplex minimization
method with multiple restarts was used to minimize the 21 free parameters in 
the CCD-to-detector plane mapping, while at each step in the simplex the 
detector plane-to-sky mapping was determined with a LU decomposition 
\citep{PR92.1} of a $\chi ^2$ minimization matrix.
The resulting best fit mappings had a positional rms difference of .06 pixels,
$0\farcs 015$, among the images of the same field and $0\farcs 55$ between the
images and the USNO-A2.0 star catalog coordinates, which is the positional 
uncertainty of this catalog.  No large scale collective offsets between the
final stellar positions and those of the USNO-A2.0 catalog were seen in any 
portion of the fields.
We also attempted the mappings using fourth to seventh order two-dimensional
polynomials for the detector plane to sky mappings, but the resulting rms
dispersions in stellar positions did not improve over the bi-cubic mapping.
A discussion of this technique in greater depth can be found elsewhere
\citep{CL02.2}.

Each CCD image was then mapped onto the common reference frame using a 
triangular method with linear interpolation which preserves surface brightness
and has been shown not to induce systematic changes in the second moments of
of objects in the case fractional pixel shift \citep{CL00.1}.  
To attempt to minimize the non-photometric conditions for the $R$-band of
A1835 and $B$ and $V$-bands on A2204,
the images were multiplied by a factor determined from comparing the fluxes of
bright but unsaturated stars to those of the image with the brightest fluxes.
For A2204 $R$-band, all twelve of the images were within $2\%$ of the brightest,
thus confirming the photometric conditions of the first night's data.  For
A1835 $R$-band, four of the twelve images were within $10\%$ of the brightest
with the remainder at 60-80$\%$ of the brightest image.  Further, three of
the twelve images had stellar FWHM significantly higher than the others, so
these were excluded from the summed image.  For the $B$ and 
$V$-band images of A2204, all of the images were within $10\%$ of the 
brightest.  The images
were then averaged using a sigma-clipping algorithm to remove cosmic rays.

\subsection{Catalog Generation and Lensing Analysis}

For object detection and photometry in the final images, we used SExtractor
\citep{BE96.1}.  For the photometric catalog, SExtractor was used
on the images with the $30\arcsec$ smoothed sky subtraction to detect objects
which had at least 5 pixels with fluxes greater than twice the signal-to-noise
of the sky after smoothing with a 3 pixel FWHM Gaussian.  The sky around each 
object was measured using a 32 pixel
thick annulus around the object, and magnitudes were measured both down to
a limiting isophote equal to the signal-to-noise of the sky and in a $2\arcsec$
radius aperture.  The best-fit Gaussian FWHM and maximum pixel flux were also 
generated to distinguish stars from galaxies.  For the A2204 $B$ and $V$-band
images, SExtractor was used in two image mode, in which objects were detected
and had the outer isophote determined in the $R$-band image, but the magnitudes
were measured in the appropriate passband.  Unless otherwise stated, all
magnitudes and fluxes are measured using the isophotal magnitudes and all
colors are from the $2\arcsec$ radius aperture magnitudes. { All magnitudes
are measured in the Vega system and determined from \citet{LA92.1} standard star
fields observed at various times during the nights.}

\begin{figure}
\centering
\resizebox{\hsize}{!}{\includegraphics{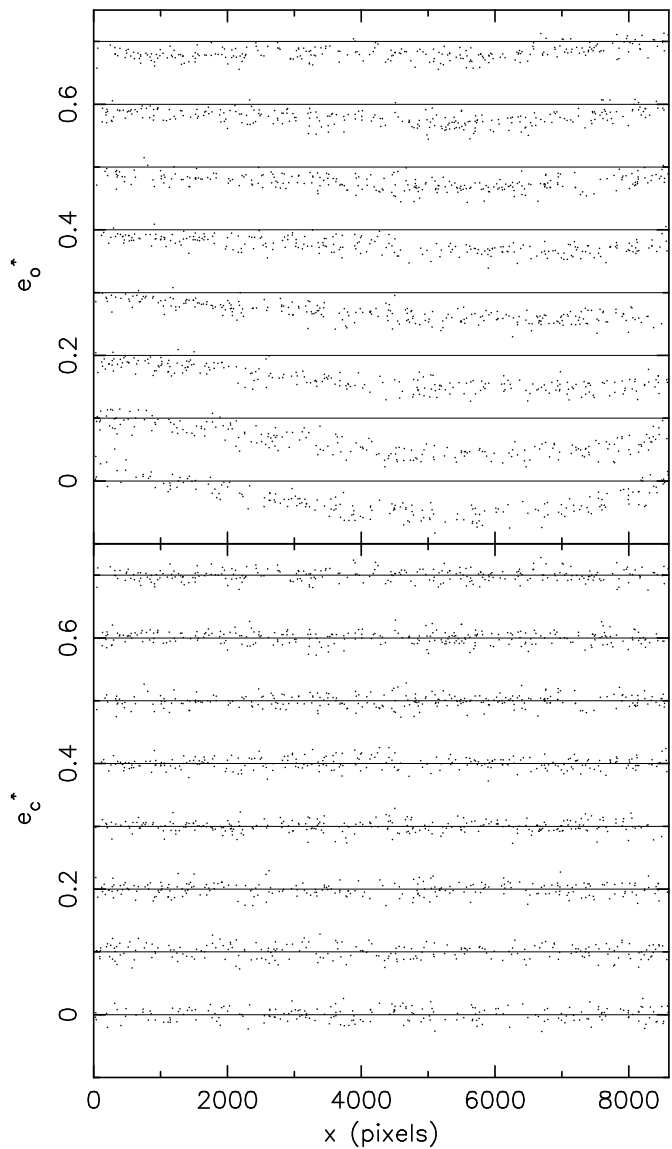}}
\caption{Above is a plot of one of the two components of the stellar PSF 
ellipticity both before (top) and
after (bottom) subtracting the best fit two-dimensional seventh order
polynomial ellipticity model for the stars.  The stars are binned into
8 vertical bins, and each bin is plotted with a 0.1 offset in ellipticity
from the others.  As can be seen, the fit adequately removes the large
scale variation in the psf ellipticity, and no sudden change in the psf
due to chip boundaries can be seen anywhere in the image.  The same is
true for the other component of the ellipticity, and is also true when
using horizontal bins and checking the ellipticity for vertical spatial
variation.  The psf has also been studied in globular cluster
fields with $\sim 20$ times greater stellar density than is present in this
image, which also fails to detect a change in the psf across the chip
boundaries \citep{CL02.2}.}
\label{psffig}
\end{figure}

For the weak lensing analysis, SExtractor was used on the $R$-band images 
with the
$7\farcs 5$ smoothed sky subtraction to detect objects which had at least
3 pixels with fluxes greater than the signal-to-noise of the sky.  The same
sky determination, Gaussian FWHM, and magnitudes were measured as given above.
The magnitudes for objects in common to both the lensing and photometric catalogs
were statistically identical for galaxies with FWHM less than $2\arcsec$,
which includes nearly all the galaxies used as background galaxies in the
weak lensing process.  The SExtractor catalog was then used with a modified
version of the IMCAT software package, written by Nick Kaiser
({\tt http://www.ifa.hawaii.edu/$\sim$kaiser/imcat}).  The objects were 
convolved with a series of Mexican-hat filters of increasing sizes in order
to determine the smoothing radius at which the objects achieved maximum 
signal-to-noise against the sky.  The radius with maximum signal-to-noise was 
then converted to a Gaussian smoothing radius (hereafter $r_\mathrm{g}$ for
the smoothing radius and $\nu$ for the signal-to-noise) and an average level
and slope of the sky were determined.    A new centroid for
each object was then computed by minimizing the first moments of the surface
brightness using a Gaussian weighting function with radius $r_\mathrm{g}$.  
At this position, a new
smoothing radius and signal-to-noise were computed for the object, and a
new centroid calculated.  This was repeated until either the new centroid
was displaced by less than $1/20$th of a pixel from the previous centroid,
in which case the object was kept in the catalog, or either the centroid
moved more than a pixel from the original SExtractor coordinates or the
signal-to-noise of the object fell below 5, in which case the object was
removed from the catalog.  For all surviving objects, a $50\%$ encircled light
radius ($r_\mathrm{h}$) was measured and the second and fourth moments of the 
surface brightness were calculated using a Gaussian weighting filter with
radius $r_\mathrm{g}$.  The second moments were then converted into
ellipticities, and the flux, second, and fourth moments were used to
calculate the shear and smear polarizability tensors ($P_\mathrm{sh}$ and
$P_\mathrm{sm}$) which define how the object reacts to an applied shear
or convolution with a small anisotropic kernel, respectively (\citealt{KA95.4},
hereafter KSB, corrections in \citealt{HO98.1}).

Unsaturated stars brighter than $R=24$ were selected by their half-light
radius and used to model the PSF.  As the half-light radius and ellipticity
of the stars varied systematically, but smoothly, over the images, the
half-light radius, ellipticity components, and the traces of the shear and smear 
polarizabilities of the stars were fit using two-dimensional seventh-order
polynomials.  It is important to note that there was no evidence for a change
in the PSF, either in ellipticity or size, across any of the regions containing
CCD boundaries.  This demonstrates both that the WFI CCDs are sufficiently
well aligned vertically to avoid changes in the PSF across chip edges and
that all of the input images had sufficiently similar PSFs such that the area
of the image which does not include contribution from any one image due to the
chip gap has the same PSF as the rest of the image.  { The spatial variation
in the PSF both before and after subtraction for the A1835 field can be seen 
in Fig.~\ref{psffig}.}

\begin{figure*}
\centering
\includegraphics[width=17cm]{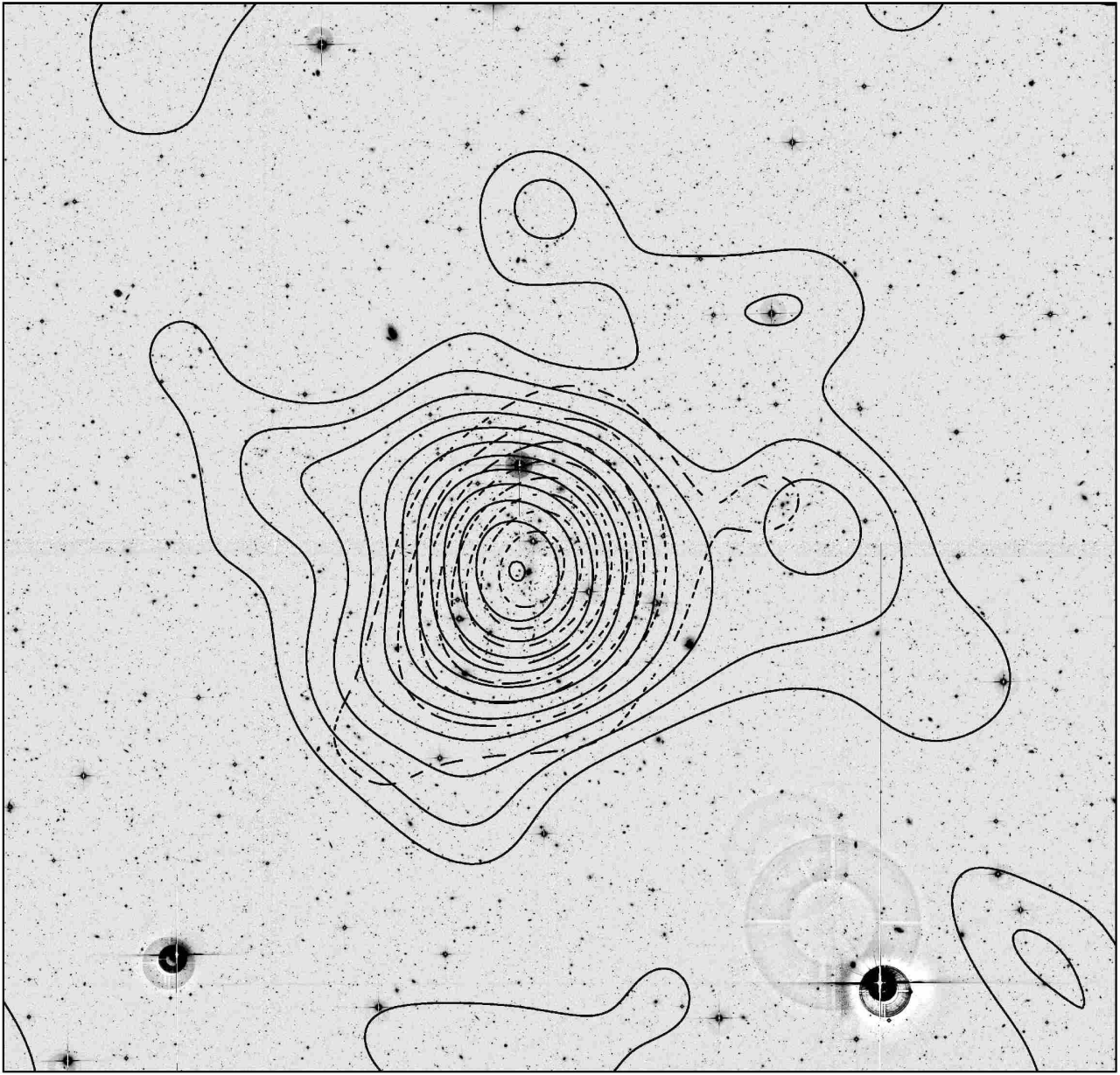}
\caption{Above is a $33\farcm 4 \times 32\farcm 1$ $R$-band image of the cluster
\object{A1835} from the Wide Field Imager on the ESO/MPG 2.2m telescope plotted using
a $\sqrt{\log }$ stretch.  The two-dimensional mass reconstruction from
the weak lensing shear signal is drawn in solid-line contours.  The input 
shear field was smoothed using a $\sigma = 1\farcm 9$ Gaussian, which is 
roughly the smoothing level of the output mass reconstruction.  Each contour 
represents an increase in $\kappa $ of 0.01 ($\sim 4.2\times 10^{13} h 
M_\odot /\mathrm{Mpc}^2$ assuming $z_\mathrm{bg}=1$) above the mean $\kappa $ at the 
edge of the field.  The dashed-line contours show the flux-weighted
distribution of galaxies with $16.78 < R < 21.78$ also smoothed by a 
$\sigma = 1\farcm 9$ Gaussian.}
\label{fig1}
\end{figure*}

The ellipticities of the galaxies were corrected using 
$\vec{e}_\mathrm{c} = \vec{e}_\mathrm{o} - (\mathrm{tr} 
\mathcal{P}_\mathrm{sm}^*)^{-1} \mathcal{P}_\mathrm{sm}
\vec{e}_\mathrm{f}^* $ (KSB) where 
$\vec{e}_\mathrm{f}^*$ is the fitted stellar ellipticity field evaluated at the position 
of the galaxy, $\vec{e}_\mathrm{o}$ is the original measured ellipticity of the galaxy,
and $\mathrm{tr} \mathcal{P}_\mathrm{sm}^*$ is the fitted trace of the stellar smear 
polarizability evaluated at the position of the galaxy.  The effects of 
circular smearing by the PSF can then be removed from the galaxies using 
$\vec{g} = (\mathcal{P}_{\gamma})^{-1} \vec{e}_\mathrm{c}$ \citep{LU97.1}
where $\mathcal{P}_{\gamma} = \mathcal{P}_\mathrm{sh} - 
\mathcal{P}_\mathrm{sm} \mathrm{tr} \mathcal{P}_\mathrm{sh}^* 
(\mathrm{tr} \mathcal{P}_\mathrm{sm}^*)^{-1}$,
for which the $\mathrm{tr} \mathcal{P}^*$ denote the fitted traces of the stellar
shear and smear polarizability evaluated at the position of each galaxy.
The resulting $\vec{g}$'s are then a direct estimate of the reduced
shear $\vec{g} = \vec{\gamma}/(1-\kappa)$, where both the shear $\vec{\gamma}$
and convergence $\kappa $, the dimensionless mass density, are second 
derivatives of the gravitational potential.  Because the measured 
$\mathcal{P}_{\gamma}$ values are greatly affected by noise, we fit
$\mathcal{P}_{\gamma}$ as a function of $r_\mathrm{g}$, and 
$\vec{e}_\mathrm{c}$.  Because the PSF size varied slightly over the image,
we divided the background galaxies into 4 bins based on the stellar 
$r_\mathrm{h}$ in their vicinity, and did the $\mathcal{P}_{\gamma}$ fitting
separately for each bin.
Simulations have shown that this technique reproduces the level of the
observed shear to better than one percent accuracy for PSFs of similar
profiles and ellipticities to those in these images \citep{VA00.1,ER01.1,BA01.1}.

\section{A1835}

Abell 1835, at $z=0.252$ \citep{AL92.1} is the most X-ray luminous
cluster in the ROSAT Brightest Cluster Sample \citep{EB98.1}.
It has been studied with both XMM \citep{PE01.1} and Chandra \citep[hereafter SA01]{SC01.1}, 
both of which demonstrate that the cluster has a relatively
cool (3-4 keV) inner core surrounded by a hotter (8-12 keV) outer envelope.
While this would suggest the presence of a massive cooling flow, a simple
model has been ruled out by the lack of emission lines in the inner region
\citep{PE01.1}.  The Chandra observations result in a best fit
NFW model with $r_\mathrm{200} = 1.28 h^{-1}$ Mpc and $c = 4.0$, for an 
$\Omega_\mathrm{m} = 1.0, \Lambda = 0$ cosmology 
(SA01).

\begin{figure*}
\centering
\includegraphics[width=17cm]{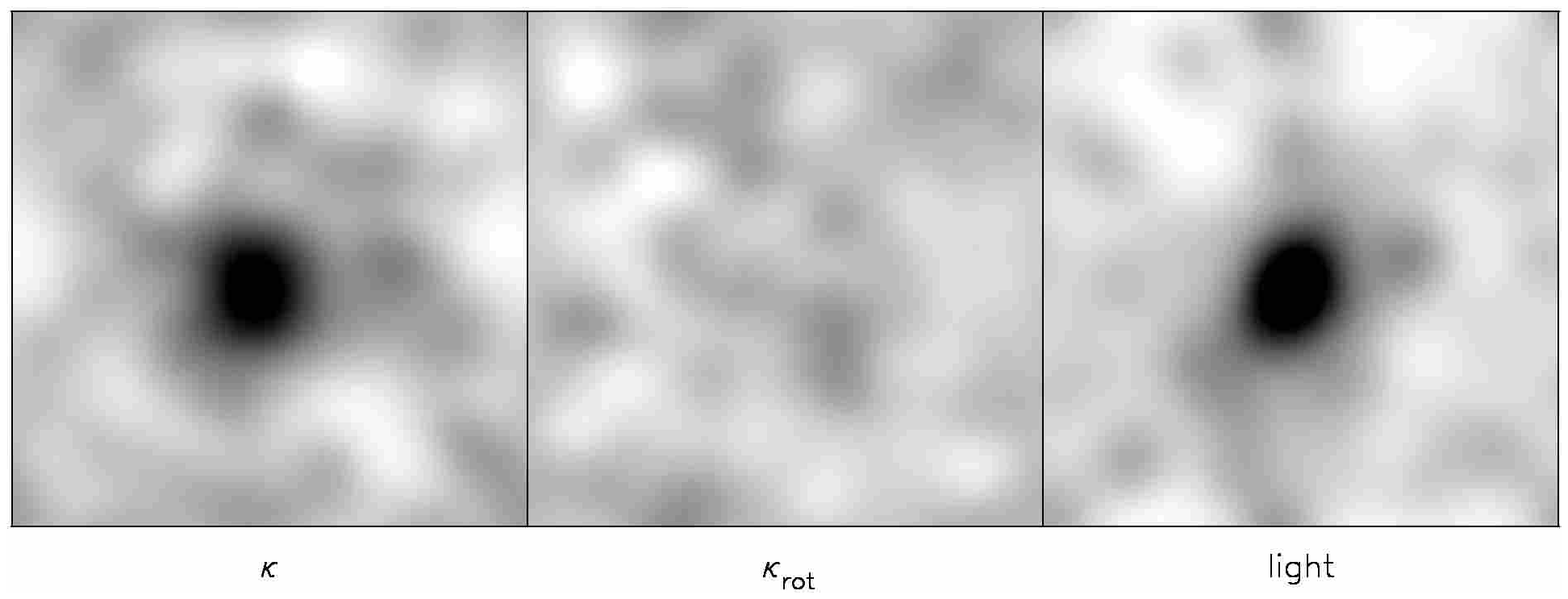}
\caption{Shown above are greyscale plots of the mass reconstruction (left)
and luminosity distribution of the $16.78 < R < 21.78$ galaxies (right) in the 
A1835 field.  Both images have been smoothed by a $\sigma = 1\farcm 9$ 
Gaussian, and are the images plotted as contours in Fig.~\ref{fig1}.
The middle image is the mass reconstruction after the shear estimators of
the background galaxies were rotated by $45^\circ$.  This provides both
a good visual estimate of the noise level in the reconstruction and a
check on any potential systematic errors from telescope optics, etc.}
\label{A1835.mrl}
\end{figure*}

A number of strong lensing arcs have been detected $\sim 30\arcsec$ from the
brightest cluster galaxy (BCG).  The three tangential and one radial arcs 
are apparently from three or four lensed sources, and can be well fit using
a NFW model consistent with the X-ray mass models (SA01).
The BCG shows evidence for recent star formation from strong optical
emission lines and UV continuum \citep{AL95.1, CR99.1} and 
850 $\mu$m emission \citep{ED99.1}.

The WFI image used in the weak lensing analysis is shown in Fig.~\ref{fig1}.  Only nine
of the twelve images taken were coadded for the final image, the remaining
three having significantly larger seeing.  The final image is $34\farcm 6
\times 33\farcm 3$ with $78\%$ having the full exposure time and the remainder
receiving lesser amounts due to the region being in a chip gap or out of the
field of view for one or more of the images.  The $1\sigma $ sky noise for the
regions with the full exposure time is 27.7 mag/arcsec$^2$.  The mean seeing
on the image is $0\farcs 72$, but this varies systematically with position by
$\pm 0\farcs 05$, with the worst seeing in the SW quadrant.  Object counts
at a $5\sigma$ detection limit in the regions containing the full exposure
time are complete to $R=24.6$ for a $2\arcsec $ radius aperture,
as measured by the point where the number counts depart from a power law.
Galactic extinction was corrected for using $E(B-V) = 0.030$ \citep{SC98.1}, 
converted to $A_R = 0.080$ \citep{CA89.1}.

\begin{figure}
\resizebox{\hsize}{!}{\includegraphics{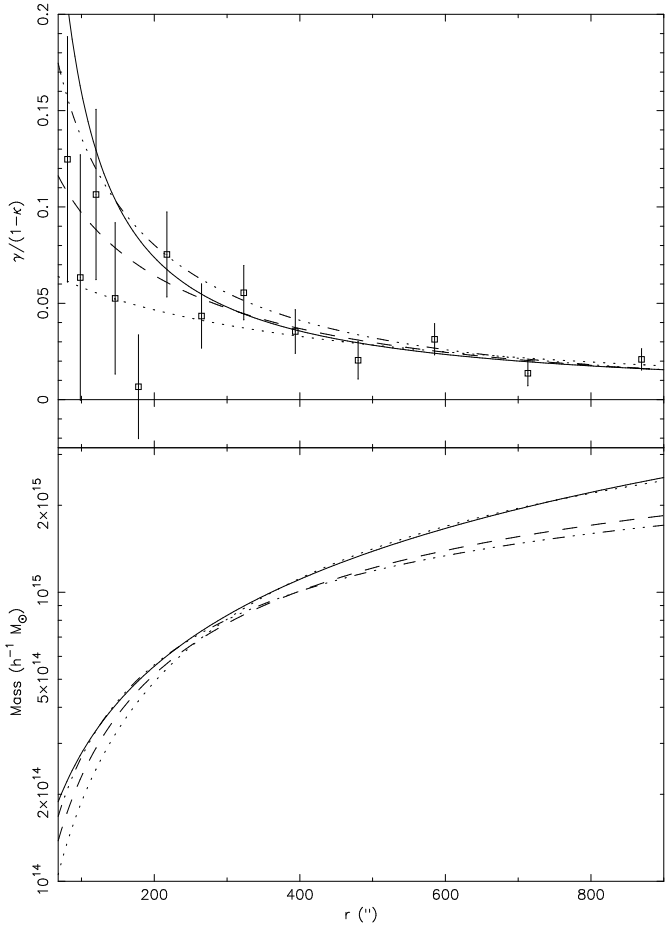}}
\caption{In the top panel above is plotted the reduced shear profile for
A1835, azimuthally averaged about the BCG, with $1\sigma$ error measurements
for each bin.  Also shown are the best fit SIS (solid line) and NFW (dashed
line) models for the fit region $3\farcm 2$ to $16\arcmin$ (800-4000 pixels),
best fit NFW model (dotted line) for the fit region $1\arcmin$ to $16\arcmin$,
and the best fit NFW model (dash-dotted line) to the Chandra data 
(SA01).  The bottom panel shows the mass profiles of the models
given in the top panel.}
\label{fig2}
\end{figure}

Using the SExtractor isophotal magnitudes, with the outer isophote at 27.70
mag/arcsec$^2$, potential cluster galaxies were selected from objects with
best fit Gaussian FWHM larger than stars and with $16.78\le R<21.78$,
where $R=16.78$ is the magnitude of the brightest galaxy of the largest
overdensity of galaxies by number in the image, and the galaxy for which
the redshift was measured by \citet{AL92.1}.  Within a $101\farcs 2$
radius aperture from the BCG (250 $h^{-1}$ kpc for $\Omega_\mathrm{m} = 1,
\Lambda = 0$ cosmology, as assumed by \citealt{BA81.1}), the overdensity is
$44^{+7}_{-5}$ galaxies
with $18.05\le R<20.05$, where $R=18.05$ is the magnitude of the third
brightest galaxy in the aperture, compared with the number density at the 
edge of the image, which corresponds to an Abell class III cluster
\citep{AB58.1}.  Using the same radius for the determination
of the second and third brightest cluster galaxies results in a Bautz-Morgan
type I \citep{BA70.1} measurement for the cluster, $(m_3-m_1)+(m_2-m_1)=2.45$.  Inside a
500 $h^{-1}$ kpc aperture ($181\farcs 4$) is a flux overdensity of
$R=13.98\pm .07$ compared to the flux density in an annulus with 
$2 h^{-1}<r<2.5 h^{-1}$ kpc.  Using a passive evolution correction { and
K-correction} on the
synthetic elliptical galaxy spectra of \citet{BR93.1} results
in a measured cluster luminosity of $L_{R,500} = (1.11\pm 0.07)\times 10^{12} h^{-2}
 L_{\odot}$.

\begin{figure}
\resizebox{\hsize}{!}{\includegraphics{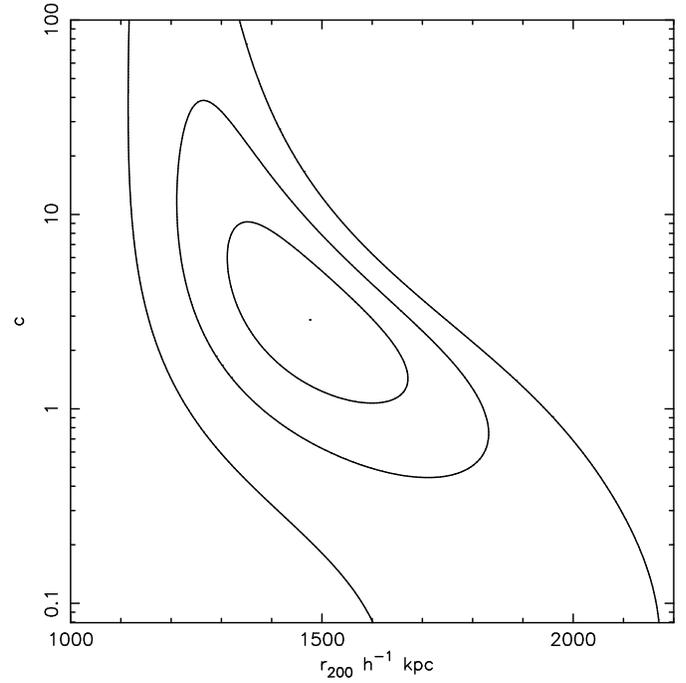}}
\caption{Shown above are the confidence contours for the NFW fit to the 
radial shear for A1835 shown in Fig.~\ref{fig2}.  The plotted contours are 
for one,
two, and three $\sigma$ confidence levels (68.3\%, 95.4\%, and 99.73\% 
respectively) as measured by the change in $\chi ^2$ from the best fit 
model.  As can be seen, because of the relatively large minimum radius
of the fit, $c$ is poorly constrained.}
\label{fig3}
\end{figure}

For the background galaxy catalog used in the weak lensing analysis, objects
which had isophotal magnitudes with $22<R<25.5$, $\nu > 7$, and $r_\mathrm{g}
> 0.32\arcsec$  (stars had $0.28\arcsec <r_\mathrm{g}<0.32\arcsec$) were 
selected from those passing all the tests given in Sect.~2.2.  
This resulted in a catalog of 15,699 objects, or 14.9 galaxies/sq arcmin, 
which can be used as a direct measure of the gravitational shear.
Shown in Fig.~\ref{fig2} is the radial shear
profile of the catalog centered on the BCG.  As can be seen, the profile
shows an increase of the shear with decreasing radius to $\sim 800$ pixels
(192\arcsec, $\sim 530 h^{-1}$ kpc), inside of which the signal falls off.
Using the shear between 800 and 4000 pixels (530--2650 $h^{-1}$ kpc) distance 
from the BCG results in best fit models of $\sigma_\mathrm{v}=1233^{+66}_{-70}$
km/s velocity dispersion for SIS
and $r_{200} = 1550 h^{-1}$ kpc, $c = 2.96$ for NFW models.  Both models have
a significance of 8.8$\sigma$, as measured by $\delta \chi^2$ from a zero
mass model, and the error contours for the NFW models can be found in 
Fig.~\ref{fig3}.
The two models are indistinguishable in terms of quality of fit.
Using the $\Omega_\mathrm{m} = 1, \Lambda = 0$ cosmology of
SA01 results in a best fit NFW model with 
$r_{200} = 1310 h^{-1}$
kpc, $c = 2.81$, which is of the same total mass but with a lower 
concentration as the Chandra fit.  The Chandra best-fit model, however, is 
within our 1$\sigma$ errors.  The mass overdensity inside a $500 h^{-1}$
kpc aperture compared to the density in a $2 h^{-1}<r<2.5 h^{-1}$ kpc
annulus gives a mass excess of $4.92^{+0.54}_{-0.55} \times 10^{14}
h^{-1} M_{\odot}$ for the SIS fit and $4.82^{+0.70}_{-0.65} \times 10^{14}
h^{-1} M_{\odot}$ for the NFW fit.  These result in mass-to-light ratios of
$M/L_R = 443^{+56}_{-57} h M_\odot /L_\odot $ and $434^{+69}_{-64} h M_\odot /L_\odot $
respectively.  The shear at the edge of the image is still measurable
at greater than $3\sigma$ significance, and the shear from all galaxies
more than $10\arcmin$ away from the BCG is detected at $5\sigma$ significance.

\begin{figure}
\resizebox{\hsize}{!}{\includegraphics{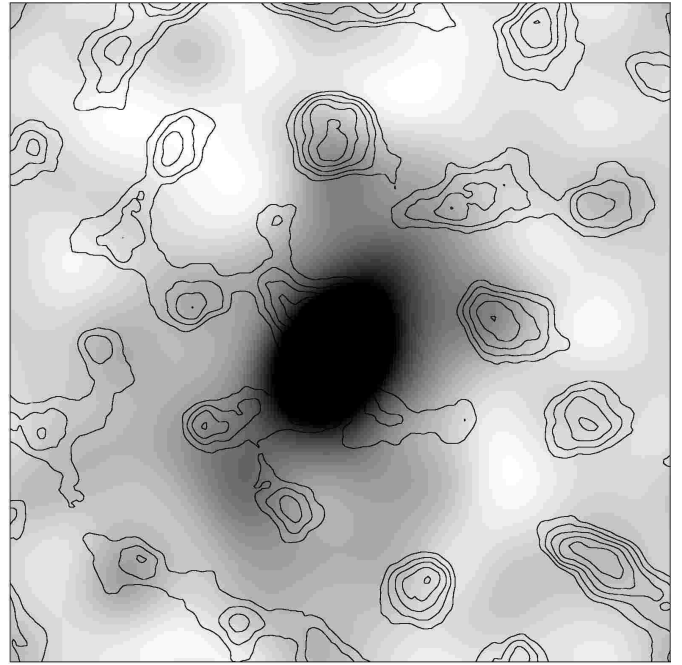}}
\caption{Shown above in greyscale is the number density of bright
($16.8<R<21$) galaxies in the A1835 field smoothed with a $1\farcm 7$
Gaussian.  In contours is the signal-to-noise measurement for a $5\arcmin $
radius mass aperture statistic measured from a catalog which had the radial
shear profile of the cluster seen in Fig.~\ref{fig2} subtracted.
Each contour is $0.5\sigma$ in significance.}
\label{galcounts1}
\end{figure}

The largest sources of possible systematic errors in these fits are the
inclusion of stars and cluster or foreground dwarf galaxies in the background
galaxy catalog and the determination of the mean redshift of the background
galaxy catalog.  The minimum size cut in the catalog, however, should have 
removed any faint stars, as well as galaxies too small to get a good 
determination of the second moment of the surface brightness.  A catalog
without the minimum size cut results in a shear estimate which is 83$\%$ of
the value with the size cut, which suggests that $\sim 20\%$ of the objects
in the full catalog are stars, in agreement with estimates from photometric
redshift catalogs of the same magnitude range \citep{FO99.1}.  The size cut removed
$\sim 24\%$ of the objects in the full catalog.  As we have only one passband
of the region, however, we cannot exclude likely cluster galaxies based on
color.  
{ Using the same magnitude cuts on the galaxies in a photometric redshift
catalog of the HDF-South \citep{FO99.1} as those used on the background
galaxy catalog results in a means lensing redshift
for the background galaxies of $z_{\mathrm{bg}} = 0.79$.  We assume that the
background galaxies lie in a sheet at this redshift for the lensing
analysis.  Contamination by cluster dwarf galaxies will lower the true mean
lensing redshift, and therefore increased the measured cluster mass, although
presumably not by a significant fraction at the edge of the field.  This
redshift determination, however, is based on only 47 galaxies in 1.4 square
arcminutes, and thus both the Poissonian error and cosmic variance are
presumably quite large.}

In the determination of the NFW $r_{200}$,
the majority of the weight in the fit comes from the outer regions of the
image.  As such, it would be heavily affected by any stellar population
remaining in the catalog, but have relatively little effect from cluster
dwarfs.  The determination of $c$ in the NFW fits, however, is strongly
dependent on the inner regions of the fit area, and as such is greatly
affected by any cluster galaxies in the catalog { (an example is given in 
the A2204 results)}.  A fit to the shear profile
from 250-4000 pixels (165-2650 $h^{-1}$ kpc) from the BCG gives $r_{200} =
1600 h^{-1}$ kpc, $c = 1.45$, and the Chandra best-fit model is outside the
$3\sigma$ contours.  If one assumes that the Chandra mass model is the true
mass model of the cluster, then these lower concentration best-fit NFW models
for the weak lensing shear profiles indicate that the fraction of cluster
galaxies in the `background' galaxy catalog is $\sim 0\%$ at distances more than
$8\arcmin$ from the cluster center, $\sim 12\%$ at distances of 
$5\arcmin -8\arcmin$, $\sim 20\%$ between $3\arcmin -5\arcmin$, and
$\sim 50\%$ at $1\arcmin -2\arcmin$.  

Overlayed in solid contours on the $R$-band image in Fig.~\ref{fig1} is the 
resulting mass map from a KS93 reconstruction \citep{KA93.1} smoothed with 
a $1\farcm 9$ Gaussian.  { Massmaps were also generated with a 
finite-field reconstruction \citep{SE96.3} using the same smoothing scale,
which show identical structures as seen in Fig.~\ref{fig1}}.  
{ Shown in Fig.~\ref{A1835.mrl} are the mass reconsruction in greyscale and
a visual estimate of the noise created by rotating all of the shear estimators
of the background galaxy catalog by $45\deg$.}
As can be seen, there is a significant detection
of the cluster mass.  The centroid of this mass is located $24\arcsec$ east of
the BCG, but this offset is not significant.  In simulations of a 
$\sigma_\mathrm{v}=1170$ km/s SIS lens with the same background galaxy
density as measured in the image, the measured centroid of the mass 
reconstructions differ from the true centroid by more than $24\arcsec$ $64\%$
of the time.  { Aside from the main cluster mass peak, several additional
mass peaks are seen in the mass reconstruction.   Significances for these
features were measured using the mass aperture statistic 
\citep{SC96.3, SC98.3} after subtracting the cluster reduced shear profile measured
in Fig.~\ref{fig2} from the shear estimators of the background galaxy catalog.
For this we used the mass aperture statistic 
\begin{equation}
M_\mathrm{ap} = 4 \sum_{i=1}^{n}\,x_i^2\,(1-x_i^2)\,g_{\mathrm{t,}i}
\end{equation}
where $x$ is the radial distance from galaxy $i$ to the aperture center
normalized by the maximum filter radius, $g_{\mathrm{t,}i}$ is the
component of the reduced shear measurement for galaxy $i$ tangential to
the center of the aperture, and the sum is taken over all galaxies within
the maximum radius of the aperture.

The significance levels measured by the mass aperture statistic for a 
$5\arcmin $ radius filter size can be seen in Fig.~\ref{galcounts1} as contour
superimposed on the number density distribution of bright ($15<R<21$)
galaxies.  There are five mass peaks which a significance greater than
$2.5\sigma$.  Four of these peaks, one $\sim 8\arcmin$ west and one 
$\sim 8\arcmin$ north of the cluster, and one each in the SW and NW corners
of the field, are spatially coincident with peaks in the bright galaxy
distribution and are seen in the mass reconstruction in Fig.~\ref{fig1}.  
With a $5\arcmin$ radius filter size, 
there are $\sim 185$ independent mass measurements in the field, and the
chance of having a 2.5$\sigma$ positive noise peak is $68\%$.  There is
only a $0.5\%$ chance of having five positive mass noise peaks at $2.5\sigma$
or higher significance.  To measure a mass for these peaks an SIS model was
fitted from $36\arcsec$ to $6\arcmin$ radius, centered on the position of 
highest signal-to-noise. The peak $\sim 8\arcmin$ west of the cluster 
was best fit by a $\sigma_\mathrm{v}=610^{+120}_{-160}$ km/s SIS with 
$2.2\sigma$ significance, and the peak $\sim 8\arcmin north$ of the cluster
was best fit with a $\sigma_\mathrm{v}=760^{+100}_{-120}$ km/s SIS with 
$3.3\sigma$ significance.  The mass measurements assume the peaks are at the
redshift of the cluster.  The other three peaks were not well fit with
a SIS model.}

\begin{figure*}
\centering
\includegraphics[width=17cm]{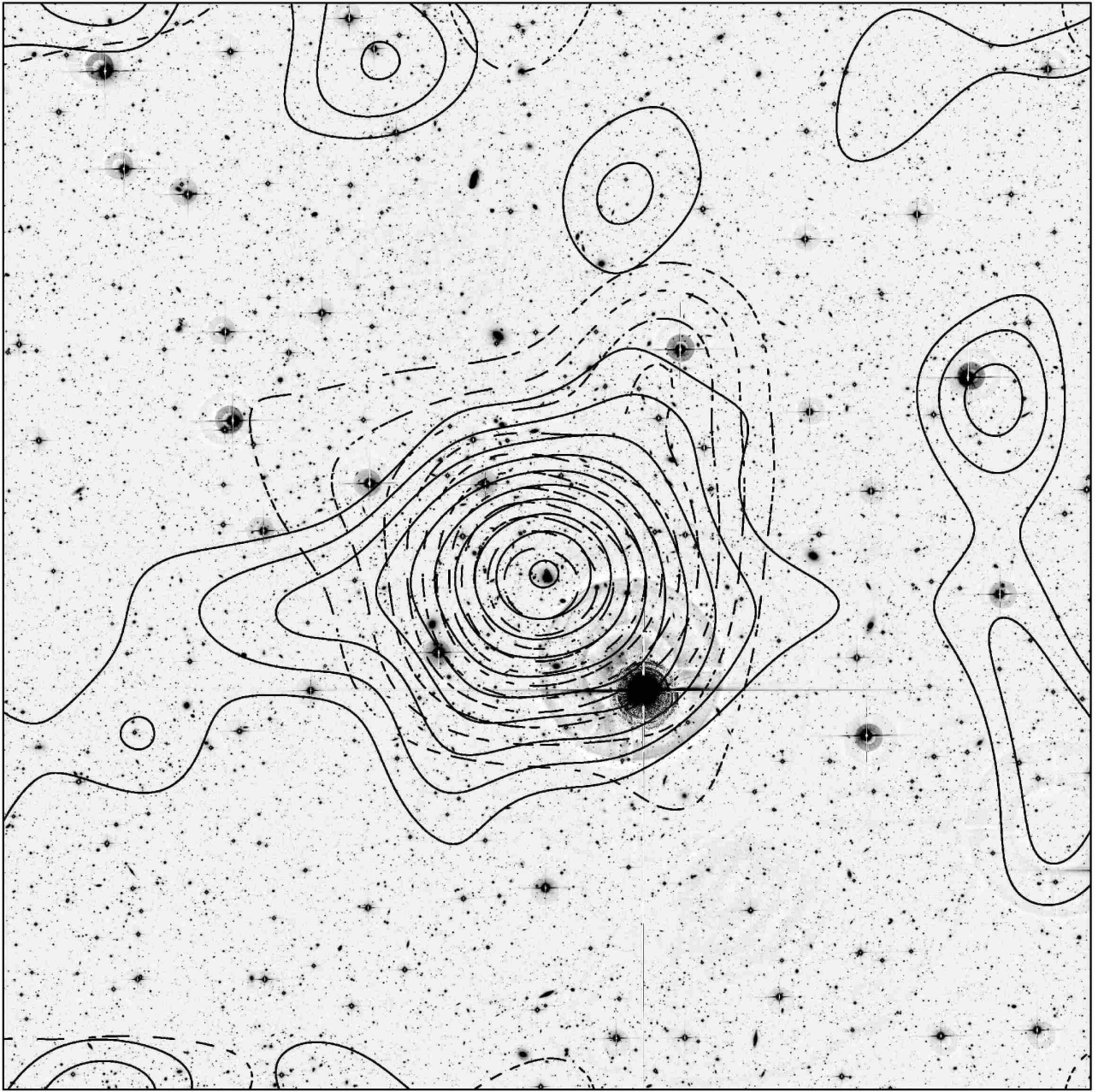}
\caption{Above is a $32\farcm 0 \times 32\farcm 0$ R-band image of the cluster
\object{A2204} from the Wide Field Imager on the ESO/MPG 2.2m telescope plotted using
a $\sqrt{\log }$ stretch.  The two-dimensional mass reconstruction from
the weak lensing shear signal is drawn in solid-line contours.  The input 
shear field was smoothed using a $\sigma = 1\farcm 9$ Gaussian, which is 
roughly the smoothing level of the output mass reconstruction.  Each contour 
represents an increase in $\kappa $ of 0.01 ($\sim 5.4\times 10^{13} 
h^{-1} M_\odot /\mathrm{Mpc}^2$ assuming $z_\mathrm{bg}=1$) above the mean $\kappa $ at the 
edge of the field.  The dashed-line contours show the flux-weighted
distribution of color-selected cluster galaxies also smoothed by a 
$\sigma = 1\farcm 9$ Gaussian.}
\label{fig4}
\end{figure*}

\section{A2204}
Abell 2204, at $z = 0.1517$ \citep{PIth}, has been a target of many
X-ray surveys \citep[eg][]{FO78.1, UL81.1, ED90.1, EB96.1, JO99.1, BO00.1}.  Using
ROSAT and ASCA data, \citet{AL00.1} obtains a best-fit model isothermal
sphere with $\sigma_\mathrm{v}=1020^{+130}_{-70}$ km/s with a small (25 kpc) core and
a massive cooling flow with $\dot{\mathrm{M}}=1007^{+98}_{-263} \mathrm{M}_\odot /\mathrm{yr}$.
Optical observations of this cluster are complicated by the relatively
low $b=33^\circ$ galactic latitude, which results in a high extinction
and stellar number count, and by the presence of a $M_V = 5.6$ star
$4\farcm 3$ southwest of the BCG.  Using the 2dF spectrograph,
\citet{PIth} has measured a velocity dispersion of $975^{+59}_{-50}$
km/s from 162 cluster galaxies.  Also along the line of sight are two
foreground groups at $z=0.058$ and 0.079 and one background group at
$z=0.292$ \citep{PIth}.   A2204 is measured to be Bautz-Morgan type II
and Abell richness class 3 \citep{AB89.1}, but the number counts may be increased
by the aforementioned foreground and background structures.

\begin{figure*}
\centering
\includegraphics[width=17cm]{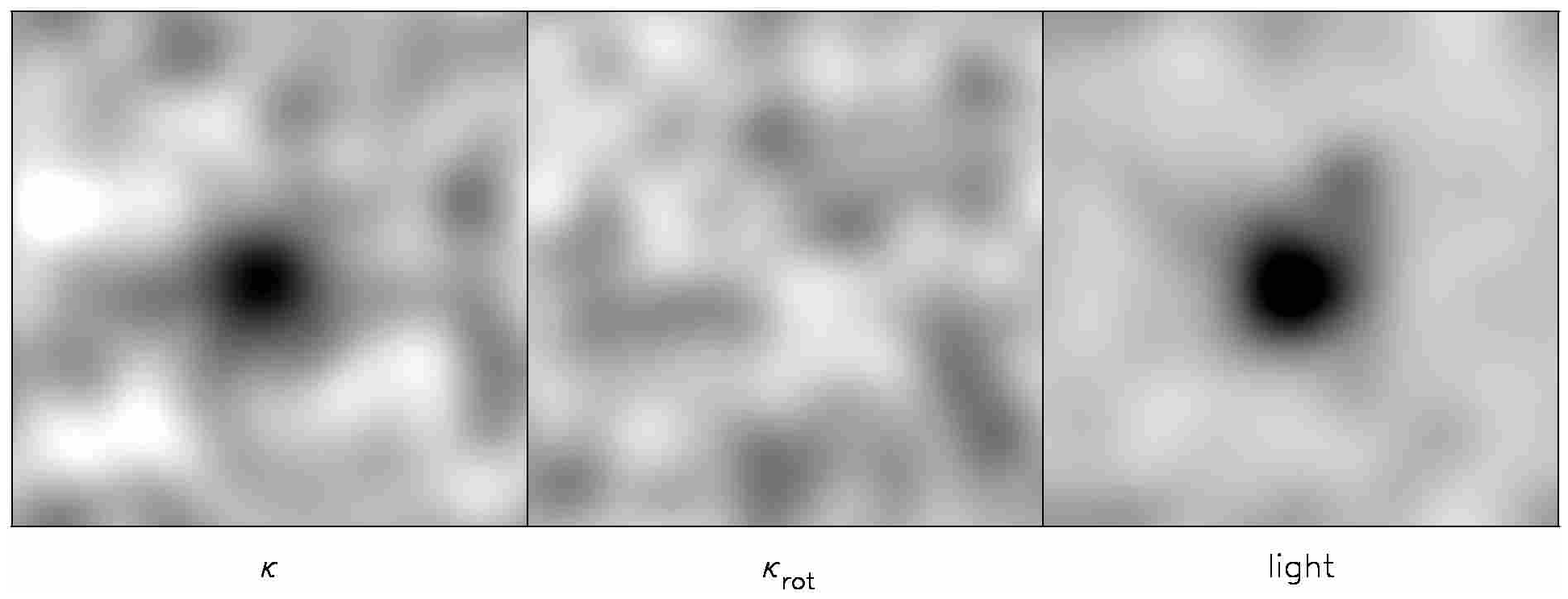}
\caption{Shown above are greyscale plots of the mass reconstruction (left)
and luminosity distribution of the color-selected cluster galaxies (right) 
in the 
A2204 field.  Both images have been smoothed by a $\sigma = 1\farcm 9$ 
Gaussian, and are the images plotted as contours in Fig.~\ref{fig4}.
The middle image is the mass reconstruction after the shear estimators of
the background galaxies were rotated by $45^\circ$.  This provides both
a good visual estimate of the noise level in the reconstruction and a
check on any potential systematic errors from telescope optics, etc.}
\label{A2204.mrl}
\end{figure*}

The WFI $R$-band image used in the lensing analysis is shown in Fig.~\ref{fig4}.  The
final image is $34\farcm 7 \times 33\farcm 4$ with $77\% $ of the area receiving
the full exposure time of 3 hours.  The $1\sigma $ sky noise for the regions 
with the
full exposure time is $28.3$ mag/arcsec$^2$.  The mean seeing on the image is
$0\farcs 79$, but this varies systematically with position by $\pm0\farcs 03$
with the worse seeing near the bottom edge.  Object counts at a $5\sigma$
detection limit in the regions containing the full exposure time are complete
to $R = 24.8$ ($R=24.55$ after correction for galactic extinction) for
$2\arcsec$ radius aperture magnitudes.  The 45 minute $V$ and 30 minute 
$B$-band images are $34\farcm 3
\times 33\farcm 6$ and $34\farcm 0 \times 33\farcm 3$ respectively.  
Only objects
in those regions common to all three images were used in the subsequent
analysis, with the exception of objects in three $\sim 20\arcsec $ wide
vertical stripes located in the chip gaps for which the $B$-band image
is blank and thus only and $V$ and $R$ magnitudes measured.  The final area
for object detection was $34\farcm 0 \times 33\farcm 1$.  Galactic
extinction was corrected for using $E(B-V) = 0.093$ \citep{SC98.1},
converted to $A_R = 0.249$, $A_V = 0.308$, and $A_B = 0.401$ \citep{CA89.1}.  

\begin{figure}
\resizebox{\hsize}{!}{\includegraphics{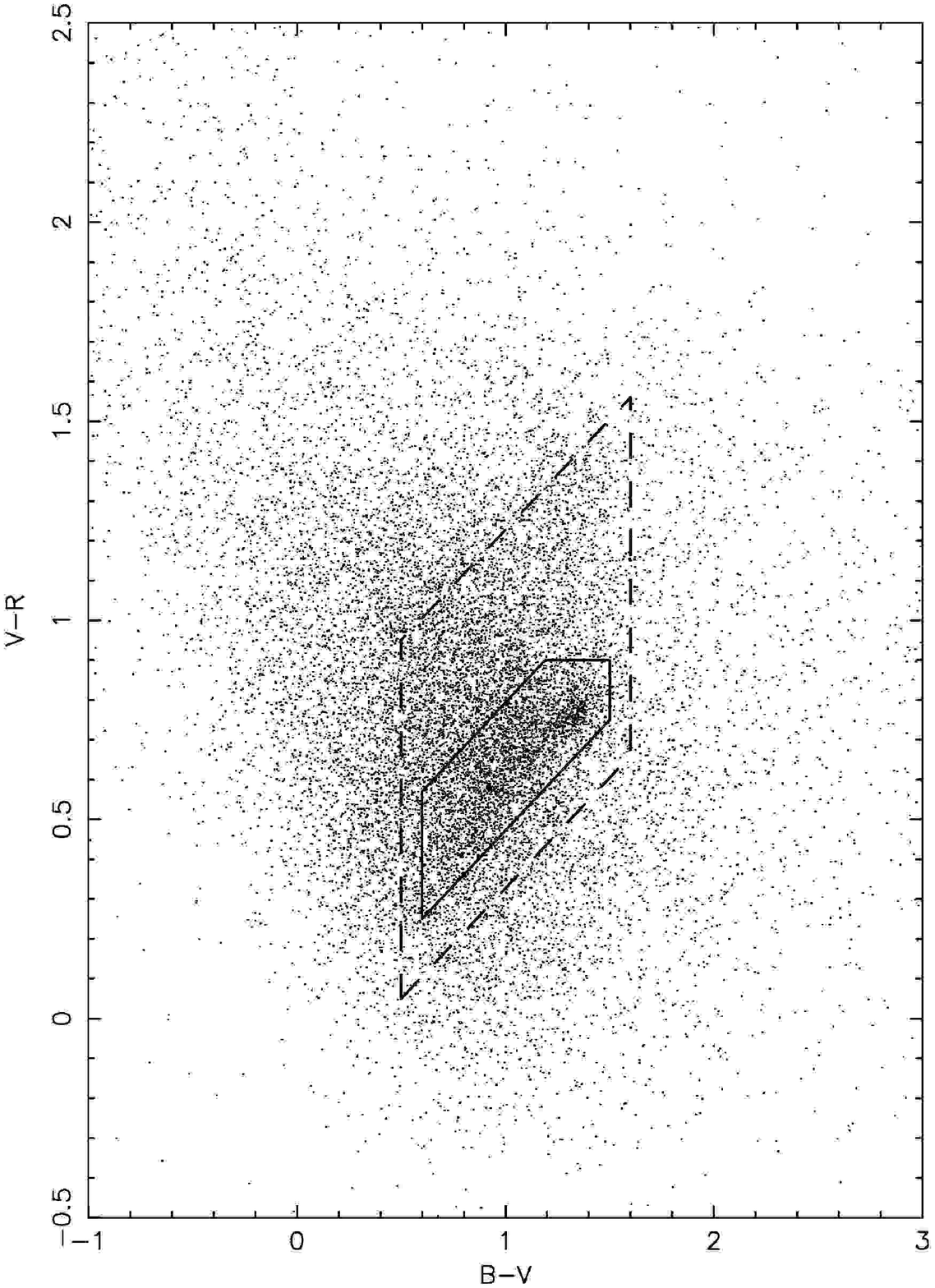}}
\caption{Above are plotted the $B-V$ and $V-R$ colors for all the galaxies
in the A2204 field with isophotal magnitudes $17<R<24$.  The area enclosed
by solid lines is the color selection for probable cluster members used
to calculate the cluster luminosity.  The galaxies in the region enclosed 
in dashed lines were excluded from the background galaxies catalog as they
have colors consistent with those of low-redshift galaxies.
}
\label{fig5}
\end{figure}

Using SExtractor isophotal magnitudes, with the outer isophote at $27.87$
mag/arcsec$^2$, for luminosity measurements and $2\arcsec $ radius aperture 
magnitudes for color measurements, potential cluster galaxies were selected
from objects with best fit Gaussian FWHM larger than stars and with 
$-0.08<(V-R)-0.56\times (B-V)<0.24$, $0.6<B-V<1.5$, and $V-R<0.9$ (see Fig.~\ref{fig5}).  
For objects in the
narrow strips not covered by the $B$-band image, potential cluster galaxies
were selected with $0.4<V-R<0.9$.  Inside a 500 $h^{-1}$ kpc aperture
(271\farcs 0) is a flux of $R=12.58\pm 0.07$.  Using a passive evolution
correction { and K-correction} on the synthetic elliptical galaxy spectra 
of \citet{BR93.1}
results in an absolute cluster luminosity of $L_{R,500} = (1.27\pm 0.08)
\times 10^{12} h^{-2} L_\odot$.  Using all of the non-stellar objects within
the 500 $h^{-1}$ kpc radius aperture with $16.2<R<22$ results in an excess flux 
of $R=12.31\pm 0.07$ compared to the flux density in a $1330 h^{-1} 
\mathrm{kpc}<r<1550 h^{-1} \mathrm{kpc}$ ($12\arcmin <r<14\arcmin $) radius 
annulus.  This results in a cluster luminosity of $L_{R,500} = (1.63\pm 0.11)
\times 10^{12} h^{-2} L_\odot$.

\begin{figure}
\resizebox{\hsize}{!}{\includegraphics{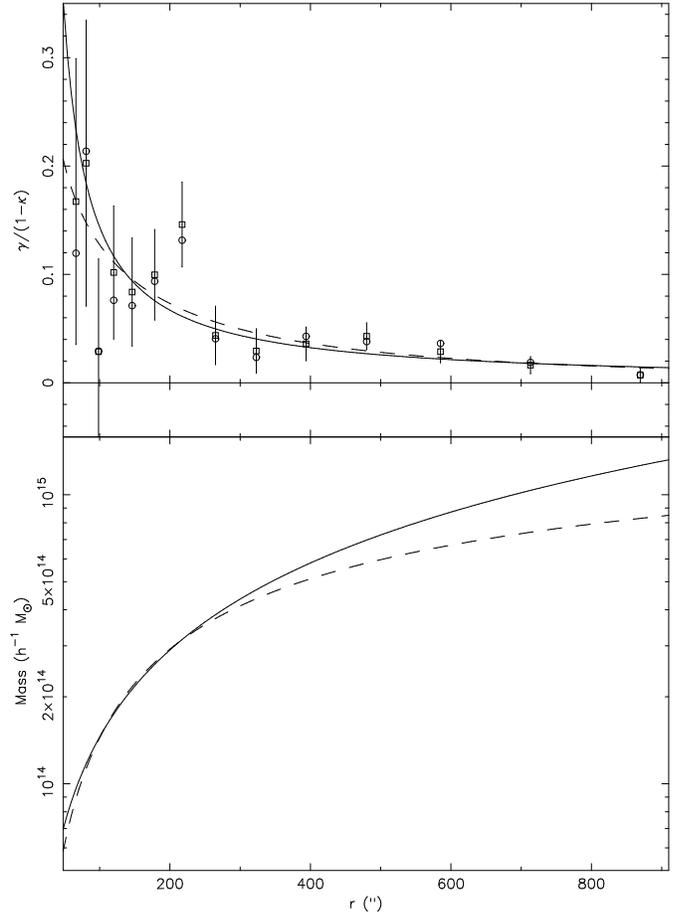}}
\caption{In the top panel above is plotted, as open squares, the reduced 
shear profile for
A2204, from a catalog of $22<R<25.8$ galaxies which had probable $z<0.5$ 
galaxies rejected by colors, radially averaged about the BCG  
with $1\sigma$ error measurements
for each bin.  Also shown are the best fit SIS (solid line) and NFW (dashed
line) models to this profile.
The bottom panel shows the mass profiles of the best fit models.
Also plotted in the top panel as open circles are the shear estimates for the
catalog of $22<R<25.8$ galaxies without the color rejection.}
\label{fig6}
\end{figure}

\begin{figure}
\resizebox{\hsize}{!}{\includegraphics{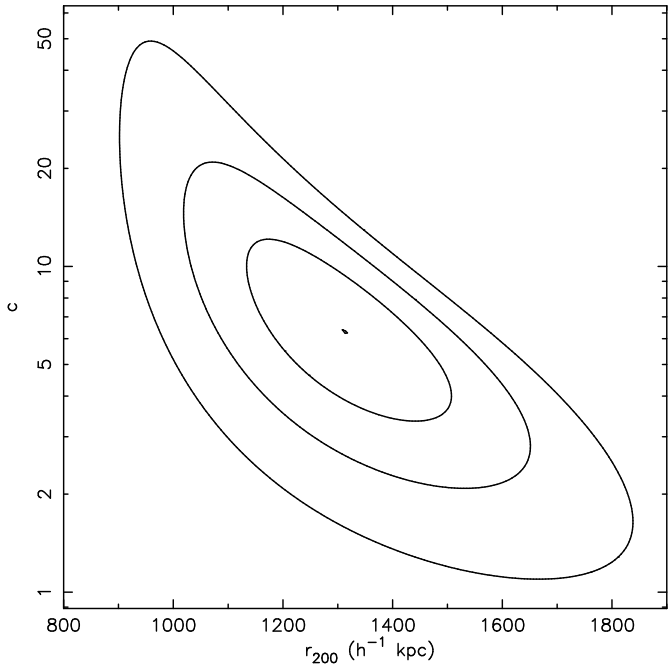}}
\caption{Shown above are the confidence contours for the NFW fit to the 
radial shear for A2204 shown in Fig.~\ref{fig6}.  The plotted contours are 
for one,
two, and three $\sigma$ confidence levels (68.3\%, 95.4\%, and 99.73\% 
respectively) as measured by the change in $\chi ^2$ from the best fit 
model.}
\label{fig7}
\end{figure}

For the background galaxy catalog used in the weak lensing analysis, objects
were selected which had isophotal magnitudes with $22<R<25.8$, $\nu > 7$, $r_\mathrm{g}
> 0.33\arcsec$, and not with colors $-0.23<(V-R)-0.56\times (B-V)<0.67$, 
$0.5<B-V<1.6$ ($0.3<V-R<1.5$ for objects in the $B$ image gaps).  The colors
were selected to exclude galaxies with $z<0.5$.  This resulted in a catalog
of 12,989 galaxies, or 12.8 galaxies/sq arcmin, after accounting for the loss
of image area due to the stellar reflection rings.  Shown in Fig.~\ref{fig6} is the
radial shear profile of the catalog centered on the BCG.  While the shear
at the edge of the image is only significant at the $1\sigma$ level, the
shear for all galaxies greater than $10\arcmin$ from the BCG is detected at
greater than $3\sigma $ significance.  Using the \citet{FO99.1} HDF-S
photometric redshift catalog, we find that the mean lensing redshift
of $z_\mathrm{bg} = 1.06$ after applying the same magnitude and color-cuts as 
mentioned above.  Using the shear
between $1\arcmin$ and $16\arcmin$ ($111 h^{-1}$--$1770 h^{-1}$ kpc) from 
the BCG results in best fit models of $1035^{+65}_{-71}$ km/s velocity
dispersion for SIS and $r_{200} = 1310 h^{-1}$ kpc, $c = 6.3$ for NFW models.
The error contours for the NFW fits is given in Fig.~\ref{fig7}.
The significances of the fits are $7.2\sigma$ and $7.1\sigma$ for the SIS and
NFW models respectively, calculated from the $\delta \chi^2$ between the best
fit and zero mass models.  The NFW model provides a marginally better fit to
the profile, with the F$_\chi$ test of additional term \citep{BE92.1}
giving an $87\% $
confidence level.  The mass contained within a $500 h^{-1}$ kpc radius for
these models are $3.89^{+0.52}_{-0.51} \times 10^{14} h^{-1} M_\odot$ for
SIS and $3.80^{+0.83}_{-0.65} \times 10^{14} h^{-1} M_\odot$ for NFW.
These result in mass-to-light ratios of $M/L_R = 306^{+46}_{-45} h M_\odot/L_\odot$ 
and $299^{+68}_{-55} h M_\odot/L_\odot$.

\begin{figure}
\resizebox{\hsize}{!}{\includegraphics{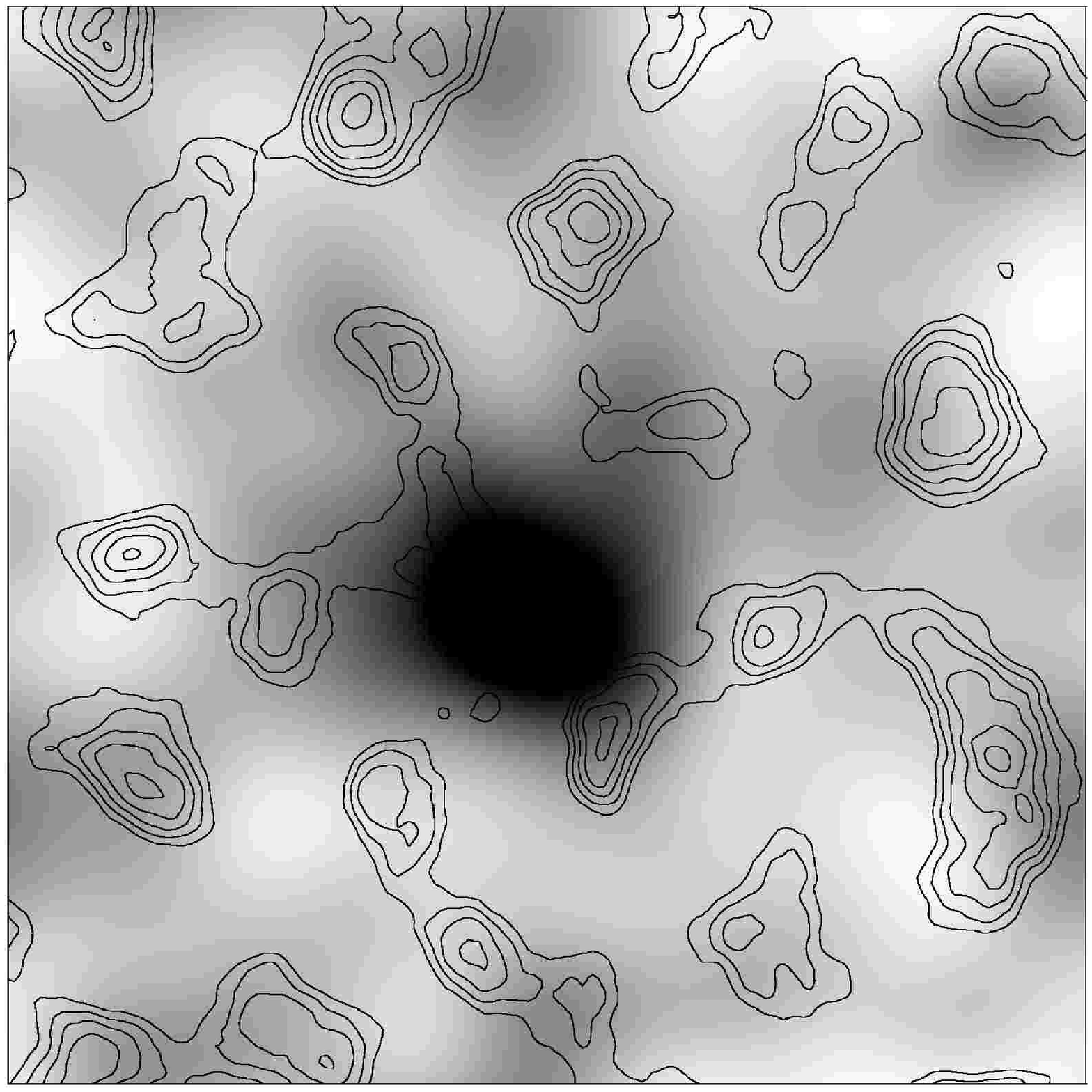}}
\caption{Shown above in greyscale is the number density of bright
($15<R<21$) galaxies in the A2204 field smoothed with a $1\farcm 7$
Gaussian.  In contours is the signal-to-noise measurement for a $5\arcmin $
radius mass aperture statistic measured from a catalog which had the radial
shear profile of the cluster seen in Fig.~\ref{fig6} subtracted.
Each contour is $0.5\sigma$ in significance.}
\label{galcounts2}
\end{figure}

Also shown in Fig.~\ref{fig6} is the shear profile of the cluster for the background
galaxy catalog without rejecting $z<0.5$ galaxies based on color.
This increases the number of galaxies in the catalog by $18.2\%$
but decreases the mean lensing redshift, as measured from the \citet{FO99.1}
photometric catalog, to $z_\mathrm{bg} = 0.82$.  Using
this catalog results in a best fit SIS model of $1054^{+60}_{-65}$ km/s
velocity dispersion and NFW model with $r_{200} = 1400 h^{-1} \mathrm{kpc},
c = 4.9$.  This is consistent with the expectations discussed in Sect.~3,
namely that because the cluster galaxies have an increasing density towards
the cluster center, the inability to exclude faint cluster galaxies from the
catalog does not significantly reduce the detected cluster mass at large
radius, but reduces the concentration in the NFW model.  While the measured
shear at large radii ($>8\arcmin$) is essentially unchanged, the shear
in the region $1\arcmin$ - $3\arcmin$ from the cluster center has been decreased by
$13\%$ from that of the color-selected catalog.  For these fits, the NFW
model provides a better fit at a $99.7\%$ confidence level, although this is mainly
due to the lower shear at smaller radius.  { The fact that the same shear 
measurement at larger radius results in a larger overall cluster mass is
due to the change in the assumed mean lensing redshift of the galaxies.  The
larger mass in this case indicates that either the field has fewer galaxies
in the $0.16<z<0.5$ region than in the HDF-S or that the color cut removed
a portion of higher-redshift galaxies not seen in the HDF-S, presumably due 
to the noise in the color measurement scattering the points into or out of
the selected region.  This would also indicate that the mean lensing redshift
for the color-selected catalog is somewhat overestimated, and that the mass
measured for the cluster is underestimated.  Simulations in which a similar
level of noise was added to the HDF-S galaxy colors and the color-selection
applied resulted in the mean lensing redshift decreasing by $\sim 0.03$ on
average, which results in an increase in the mass models of $\sim 1\%$.}

The KS93 mass reconstructions, smoothed by a $1\farcm9$ Gaussian,
is shown in solid contours on the $R$-band image in Fig.~\ref{fig4}.  
{ Shown in Fig.~\ref{A2204.mrl} are the mass reconsruction in greyscale and
a visual estimate of the noise created by rotating all of the shear estimators
of the background galaxy catalog by $45\deg$.}
As can be seen,
there is a significant detection of cluster mass, and the peak of the mass
distribution is coincident with the BCG.  In addition to the main cluster peak,
several additional structures are see in the mass map.  Both the mass peak
$12\arcmin $ NNW of the cluster center and the filamentary-like structure 
extending to the east in the mass reconstruction are also found as 
overdensities of faint galaxies with colors consistent with the cluster
galaxies.  { A smoothed galaxy number density of all galaxies with 
$15<R<21$ is shown in Fig.~\ref{galcounts2} along with signal-to-noise 
contours for a $5\arcmin $ radius mass aperture statistic calculated on
the shear estimators of the background galaxy catalog after subtracting the
cluster radial shear profile shown in Fig.~\ref{fig6}.}

Using the mass aperture statistic with a $5\arcmin $ radius,
the NNW peak has a significance of $2.8\sigma$, while the filamentary
structure is associated with 2 peaks, a $2.6\sigma$ significance 
peak $\sim 12\arcmin$ due east of the cluster and a $2.8\sigma$ peak
$\sim 5\farcm5$ south of there.  The fact that the structure which appears
filamentary in Fig.~\ref{fig4} is broken into two peaks by the mass aperture 
statistic is a result of the compensated filter used in the measurement.  
None of the three peaks is fit well by a SIS model.  { There are eight
mass peaks with a significance of at least $2.5\sigma$ when using the
$5\arcmin$ radius mass aperture statistic.  The expected number of positive
$2.5\sigma$ mass peaks due to noise is $\sim 1.2$, with a $\sim32\%$ of
having at least two such peaks.  The chance that all 8 peaks are due only
to noise is $\sim 0.002\%$.  In general, these peaks are spatially coincident
with overdensities in the bright galaxy distribution, with offsets of 
$\sim 1\arcmin$ expected due to noise.  Several of the mass peaks, however,
have maximum signal-to-noises given by larger filter radii.}

The mass peak $\sim 13\arcmin$ WNW of the cluster has a $3.1\sigma$ 
significance as measured from a $7\arcmin $ radius mass aperture.
The peak is spatially coincident with an overdensity of faint galaxies
with $V-R$ colors $\sim 0.2$ magnitudes redder than the cluster sequence,
which would imply that the peak is associated with a group of galaxies at 
$z\sim 0.3$.  Placing this peak at $z\sim 0.3$ results in a best fit SIS 
model of the cluster-shear subtracted background galaxy catalog from 
$36\arcsec $ to $7\arcmin$ of $755^{+95}_{-115}$ km/s velocity dispersion,
and is distinguishable from a zero mass model at $3.4\sigma$, as measured 
by the $\delta \chi^2$.  While the colors of the galaxies near this peak
are consistent with the $z=0.292$ structure in \citet{PIth}, the best-fit
lensing mass is ten times that inferred from the galaxy velocity dispersions.
{
There are two additional structures at the WSW and NNE edges of the image.
Both of the structures are significant at $3.5\sigma $ using a $7\arcmin$
radius mass aperture, while the chances of having $3.5\sigma$ noisepeak
from a $7\arcmin$ radius mass aperture statistic in this image is only $2.3\%$.
Both peaks are near galaxy overdensities which are redder than the cluster
galaxies, suggesting redshifts of $\sim 0.3-0.4$.}

\section{Discussion and conclusions}

We have detected a shear profile for the clusters A1835 and A2204 over a
$1\arcmin < r < 16\arcmin$ range of radii.  Fitting a SIS model to these
shear profiles gives best fit velocity dispersions of 1180 km/s for A1835 and
1040 km/s for A2204, in agreement with the mass measurements from X-ray
observations, and significances of 8.8$\sigma$ and 7.2$\sigma$ from zero-mass
models, respectively.  A2204 has a best fit NFW profile of 
$r_{200} = 1310 h^{-1} \mathrm{kpc}, c = 6.3$ which is a better fit to the 
data at $87\%$ confidence.  A1835 has a best fit NFW profile of
$r_{200} = 1480 h^{-1} \mathrm{kpc}, c = 2.86$ which is statistically
indistinguishable in quality of fit from the SIS model.  For both clusters
we detect the shear between $10\arcmin$ and $16\arcmin$ from the BCG at 
high significance, which suggests that one will still measure a shear signal
from such clusters with even larger format cameras.

\begin{figure}
\resizebox{\hsize}{!}{\includegraphics{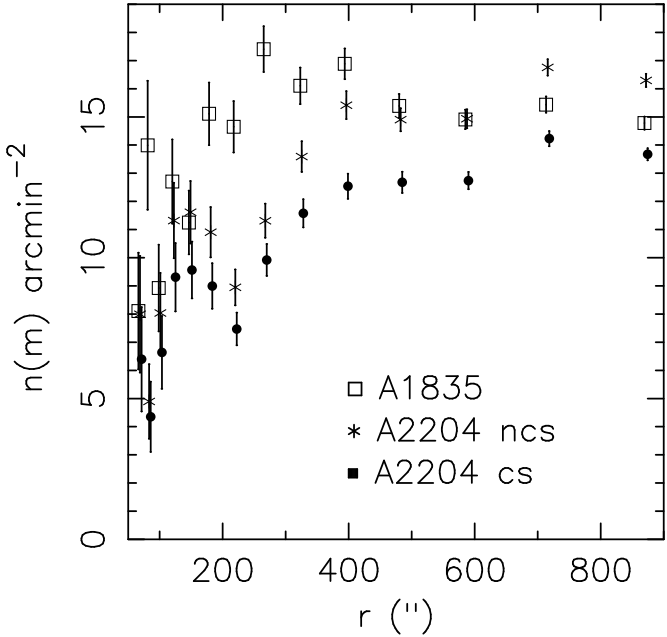}}
\caption{Shown above is the galaxy number density in radial bins around
the BCG of each cluster, after correction for loss of observable area due
to bright objects, stellar reflection rings, etc.  
The data for A2204 is shown for all the faint
galaxies (ncs) and only the galaxies surviving the rejection of probable
low-z galaxies by color (cs).  All three sets of radial bins have the same
radial range per bin but are offset from each other in the figure for greater
clarity.  The A1835 data points are plotted at the true geometrical mean
of each bin.  The error bars give the $1\sigma$ Poissonian noise for the bins,
but are underestimates of the true error as they do not allow the galaxies
to be correlated.  The depletion in galaxy number counts at small radius is
caused primarily by the rejection of galaxies with near neighbors from the
'background' galaxy catalog and not by gravitational lensing.  
As the cluster core has the highest number
density of objects in the field, the fraction of galaxies rejected by this
criteria increases in this region.}
\label{bgfig}
\end{figure}

While the total mass measured for A1835 is in agreement with the Chandra X-ray
observations of SA01, the concentration of mass towards the
center of the cluster is much smaller.  We believe this is due primarily
to contamination of the `background' galaxy catalog used to measure the shear
with cluster dwarf galaxies.  This would result in the shear in the inner
regions of the cluster being reduced due to the presence of a population of zero
shear objects, but the shear at large radii being virtually unchanged due to
the decrease in number density of the cluster dwarfs with increasing distance
from the cluster.  We measured this effect in the A2204 data by comparing a
catalog of galaxies which had probable cluster members rejected by color with
a catalog without any color rejection.  The color-selected catalog did have a
higher shear in the inner few arcminutes of the cluster radial shear profile
while having essentially the same shear at larger radius.  The amount of the
increase in shear by rejecting cluster dwarf galaxies in A2204 would not be
sufficient to bring the X-ray and weak lensing mass concentrations of A1835
into agreement.  Using the cluster dwarf number counts of Coma \citep{TR98.2}
we find from the observed magnitude limits that, given the same cluster
richness, A2204, at $z=0.152$, would have 2.5 times more cluster dwarfs in 
the `background' galaxy catalog than A1835, at $z=0.252$, but spread over 
a 2.2 times greater area.  Thus, the actual contamination of the
A1835 signal should be less than that observed in A2204, but in order to
reconcile the shear profile with the X-ray observations, we need the
contamination to be three times larger than that in A2204.

{ The galaxy density profiles in the background galaxy catalogs for the two
clusters can be seen in Fig.~\ref{bgfig}.  The rejection of low-z galaxies
by color in the A2204 data results in a decrease in the number density of
galaxies used for the weak lensing analysis of $\sim 17\%$ at large radius
and $\sim 25\%$ at small radius.  As can be seen, while A1835 has a similar
number density of galaxies to the catalog without color selection for A2204
at large radius, it has a much larger galaxy density at small radius.    
There are three possible explanations for this increase
in galaxy density in A1835 at small radius.  The first is an increased number
of cluster dwarf galaxies in the core of A1835 compared to A2204.  
The second is that
there is either an overdensity of background galaxies near A1835 or an
underdensity near A2204.  The first explanation would result in an
increase in the cluster dwarf fraction at small radius for A1835, and 
thus a greater decrease in the measured mass concentration than is seen
in the non-color selected catalog for A2204, while the opposite would
be true for the second explanation.

The third possible explanation of the increased galaxy density is that the
decrease in the number density of galaxies in A2204 with decreasing radius
 is mostly due to the rejection of galaxies from the background galaxy
catalog which have near neighbors and the increased bright galaxy density
at the cluster core.  A catalog of faint galaxies without this near neighbor
rejection actually shows a slight rise in galaxy density at small radius
in A2204, consistent with the small increase in color selected galaxies
with decreasing radius discussed above.  Because A1835 is at higher
redshift than A2204, however, the cluster core has a much smaller angular
size.  As a result, the increase in rejection of galaxies from the 
background galaxy catalog should occur at a smaller angular radius in A1835
than in A2204, which is what appears to be occurring in Fig.~\ref{bgfig}.
The fact that the number density profile of galaxies without the near neighbor
rejection is similar for the two clusters suggests that this third
explanation is the most likely, and thus the lensing mass profile does,
in fact, have a smaller concentration than that measured from the X-ray data.}

One possibility for the lower concentration in the weak lensing mass model
than in the X-ray mass model would be an elliptical cluster core.  The X-ray
NFW profile was measured by using a spherical 3-dimensional model which was
compared with a deprojected spectrum of the Chandra data.
The weak lensing profile, however, is created by first integrating the
mass along the line of sight, then fitting with a circularly symmetric model.
In the case of a prolate cluster core, if the major axis of the core is
lying in the plane of the sky then the best-fit weak lensing profile will
have a smaller concentration than that of the X-ray model due to the fact that
in the weak lensing analysis the mass is projected into a plane before being
circularly averaged, while in the X-ray analysis the mass is spherically
averaged.  Conversely, if
the major axis is lying along the line of sight, the weak lensing profile will
have a higher concentration than the X-ray model.  In studies of a CDM N-body
cluster, \citet{KI01.1} found that the best fit 2-dimensional NFW
concentration parameter among different projection axes could vary 
by $\sim 20\%$.

In addition to the primary cluster mass peak, several additional peaks were found
in both clusters at $\sim 3-4\sigma$ significance using 
$4\arcmin $-$7\arcmin $ radius mass aperture statistics.  Only two of the
six peaks are not spatially coincident with galaxy overdensities, but both
are located at the edge of the image, and thus could be associated with
structures immediately outside the field of view.  In addition, the mass
reconstructions of both clusters have filamentary-like structures extending
from the cluster cores.  These structures are also seen in faint galaxy
distributions, and part of the structure in A1835 is spatially coincident
with an X-ray source.  

{ Finally, we note that all of the masses quoted herein have assumed that the
background galaxies have the redshift distribution given by the \citet{FO99.1}
HDF-S photometric redshift catalog.  The appropriate magnitude and color cuts
were made to this catalog, and the mean lensing redshift for the sample
was calculated for each cluster.}  While the
mean lensing redshift of the background galaxies does not affect the 
significance
of the shear signals, it does affect both the total mass of the clusters and
the concentration parameter of the best fit NFW profile.  An overestimate
of the mean background galaxy redshift would result in both the best-fit mass
and concentration value for a cluster being lower than the true value.
Because both A1835 and A2204 are at relatively low redshifts, the measured
mass changes slowly with the mean lensing reshift.  {For the A2204 color
selected catalog, the
background galaxy mean lensing redshift would need to shift from the measured
$\bar{z}_\mathrm{bg} = 1.06$ to $\bar{z}_\mathrm{bg}\sim 0.6$ or
$\bar{z}_\mathrm{bg}\sim 4$ in order for the systematic error from the
redshift estimate to equal the level of the random error from the intrinsic 
galaxy ellipticities and mass sheet degeneracy.  For the A1835 catalog,
the background galaxy mean lensing redshift would need to shift from the
measured $\bar{z}_\mathrm{bg} = 0.79$ to $\bar{z}_\mathrm{bg}\sim 0.62$ or
$\bar{z}_\mathrm{bg}\sim 1.1$ for the systematic error in the assumed mean
lensing redshift to equate to the random error from the background galaxy
distribution.}

\begin{acknowledgements}
We wish to thank Lindsay King, Jean-Paul Kneib, Robert Schmidt,
and Ian Smail for help and useful discussions.
This work was supported by the TMR Network
``Gravitational Lensing: New Constraints on Cosmology and the
Distribution of Dark Matter'' of the EC under contract
No. ERBFMRX-CT97-0172 and a grant from the Deutsche Forschungsgemeinschaft.

\end{acknowledgements}

\bibliographystyle{apj}
\bibliography{H3633}

\end{document}